\documentclass[fleqn,usenatbib]{mnras}

\usepackage{newtxtext,newtxmath}

\usepackage[T1]{fontenc}
\usepackage{comment}
\usepackage{graphicx}
\usepackage{dcolumn}
\usepackage{bm}
\usepackage{textcomp}
\usepackage{amsmath}
\usepackage{verbatim}
\usepackage{comment}
\usepackage{lipsum}
\usepackage{natbib}
\usepackage{gensymb}
\usepackage{xcolor}
\usepackage{color}
\usepackage{soul}
\usepackage{cancel}
\usepackage{ulem}

\DeclareRobustCommand{\VAN}[3]{#2}
\let\VANthebibliography\thebibliography
\def\thebibliography{\DeclareRobustCommand{\VAN}[3]{##3}\VANthebibliography}

\usepackage{graphicx}	
\usepackage{amsmath}	

\title[]{Coupling of neutrino beam-driven MHD waves and resonant instabilities in rotating magnetoplasmas with neutrino two-flavor oscillations}

\author[]{
Jyoti Turi,\thanks{E-mail: jyotituri.maths@gmail.com}
Amar P. Misra, \thanks{E-mail: apmisra@visva-bharati.ac.in}
\\
Department of Mathematics, Siksha Bhavana, Visva-Bharati University, Santiniketan-731 235, West Bengal, India\\
}

\pubyear{\the\year{}}

\begin{document}
\label{firstpage}
\pagerange{\pageref{firstpage}--\pageref{lastpage}}
\maketitle

\begin{abstract}
We present an analysis of neutrino-driven magnetohydrodynamic (MHD) waves and instabilities in a rotating magnetoplasma with weak neutrino interactions. We show that neutrino-driven shear Alfv{\'e}n and oblique magnetosonic waves can be coupled by the Coriolis force, forming new wave modes affected by this force, as well as neutrino beam and two neutrino flavor oscillations. Our work extends previous theories by demonstrating that shear Alfv{\'e}n waves are influenced by neutrino effects and by identifying instabilities resulting from resonant interactions with both a streaming neutrino beam and flavor oscillations. We find that the Coriolis force, plasma density, and magnetic field strength significantly affect the profiles of instability growth rates.   Such a growth rate for magnetosonic waves appears much higher than the Alfv{\'e}n wave, implying that magnetosonic waves provide a superior mechanism for energy extraction from the neutrino beam. For typical parameters relevant to the protoneutron star surface, the instability time for magnetosonic waves may vary in the range 0.09-0.14 s, which is within the predicted time of the neutrino-driven explosion (0.3 s after bounce) reported in the recent three-dimensional MHD simulations of core-collapse supernovae. Our findings may shed new light on the physical mechanisms underlying core-collapse supernovae. 
\end{abstract}




\section{INTRODUCTION} \label{Sec-Intro}
 Neutrinos are elusive particles usually emitted during high explosions in the cores of massive stars. They weakly interact with matter but play a significant role in various astrophysical situations. For example, neutrinos appear in supernova explosions near the interior of protoneutron stars, and in the cooling of white dwarfs and neutron stars \cite{raffelt1996stars,haas2016neutrino,adams1963neutrino}.
The Sun is considered one of the prominent sources of neutrinos, which a simple fusion reaction can produce. The latter occurs when two protons form a deuterium nucleus. However, other complex mechanisms also stimulate neutrino generation, such as cosmic rays interacting with nuclei in Earth's atmosphere.
The impact of neutrinos becomes relevant when massive stars, with masses of roughly $8 M\odot-25 M\odot$, end their life via core collapse. This process forms black holes and other neutron stars, as well as type II supernova explosions \cite{sharma2021modes}. These emitted neutrinos can become so intense that they disturb the surrounding background matter. Previous studies reported that the neutrino beam during a supernova burst imparts energy and momentum to magnetized, dense plasmas around the core. This interaction triggers stalled supernova shocks \cite{serbeto2002fluid,aftab2022neutrino}.
Moreover, as it propagates through a dense plasma, an intense neutrino beam can produce an effective electric charge. This charge can transfer energy and momentum into the plasma medium via neutrino-electron interactions \cite{bingham1996nonlinear,serbeto2002fluid}.
\par 
Typically, we describe the neutrino-electron interactions via the standard electroweak models, which unify the weak force and electrostatic forces \cite{salam1959weak}. Nowadays, neutrino-plasma coupling has gained great attention among the plasma physics as well as particle physics community, as the role of the neutrino beam is becoming unavoidable in various dense astrophysical and cosmological processes, such as supernova explosions and neutron star cooling \cite{bethe1985revival,bethe1990supernova,janka2007theory}.
\par 
The magnetohydrodynamic (MHD) theory is useful for studying complex phenomena in astrophysical plasmas with strong magnetic fields. However, to understand the role of neutrino particles in magnetized plasmas, Haas \textit{et al.} \cite{haas2016neutrino} proposed a fluid model that incorporated interactions of neutrino and plasma constituents. This model is known as the neutrino MHD (NMHD) model. In that model, linear magnetosonic wave propagation was considered in a typical real scenario for a type II supernova, within a particular geometry. They demonstrated the destabilizing impact of neutrino beams on magnetosonic waves through resonant interactions. Later, the authors generalized to consider oblique magnetosonic wave propagation in an arbitrary direction to the magnetic field \cite{haas2017instabilities}. 
\par 
Although realistic neutrino winds can have a finite angular and energy spread for which the kinetic description may be necessary \cite{horn1984}, a fluid model for a neutrino beam is justified when collective interactions (neutrino-neutrino and neutrino-matter) are frequent enough for the system to reach local thermodynamic equilibrium, allowing it to behave as a continuous medium. The conclusions from this model are generally not robust to significant kinetic effects, such as large angular spread and velocity-space dispersion, as these effects violate the core assumptions of the fluid approximation and require a more complete kinetic (Boltzmann) treatment for accuracy. 
The fluid model is generally more robust at very large scales (much larger than the free-streaming scale, $L_0$), where neutrino velocities get averaged into a single fluid velocity, and the pressure effects of free streaming get modeled with effective, scale-dependent sound-speed and shear-stress terms. On scales smaller than $L_0$, the fluid approximation fails because it cannot capture the collisionless damping of structures caused by this dispersion.  
However, in extreme astrophysical environments, such as those in supernovae, the density of neutrinos and background matter (electrons, protons) is so high that, despite the intrinsic weakness of the weak force, the mean free path ($\lambda$) of neutrinos becomes significantly smaller than the characteristic length scale ($L$) of the system ($\lambda\ll L$). In this regime, neutrinos get trapped in the interior of protoneutron stars (PNSs), making the neutrino scattering rate high enough that they diffuse out rather than stream freely, and thus act collectively, allowing us to describe them using NMHD \cite{yamamoto2016}. 
\par 
 Significant advancements have been made in neutrino-driven hydrodynamic waves and instabilities over the past few decades. To mention a few, an investigation by Chiueh \cite{chiueh1993anomalous} revealed that coupling of neutrino beam with ion-acoustic waves leads to neutrino-driven instability with a growth rate of order of the Fermi coupling constant, $G_F$, which dominates the viscous damping of acoustic waves. Bingham \textit{et al.} \cite{bingham1994collective}
addressed the complex interactions of neutrinos and dense plasmas for the typical environment of the supernova core. The study showed that neutrino beams couple to collective plasma oscillations and convert neutrino energy into Langmuir waves, heating the plasma electrons via collisional damping. There also appeared the possibility of the onset of two-stream instability \cite{shukla1999physics}. Serbeto \textit{et al.} \cite{serbeto2002fluid} studied intense neutrino beam modified ion-acoustic waves in supernova II environments and reported that the neutrino-plasma interactions can transfer energy and momentum in the media and enhance stalled supernova shocks. Several authors have reported the collective interactions of neutrino beams and plasmas in various astrophysical contexts (See e.g., Refs. \cite{bingham1996nonlinear,serbeto1999solitons,serbeto2002neutrino,prajapati2017influence}).  
\par 
Depending on the propagating medium, a two-way periodic transformation of neutrinos from one state to another occurs, referred to as neutrino flavor oscillations \cite{mikheev1986neutrino,mikheev1987resonance,smirnov2005msw}. Neutrino interactions initiate resonant coupling among various flavor states, generating an effective induced neutrino charge in the presence of electric and magnetic fields. The fields then stimulate the collection plasma process, increasing the collisional cross section \cite{wolfenstein2018neutrino}. In this context, Mendonca and Haas \cite{mendoncca2013neutrino} proposed a model with neutrino flavor and plasma oscillations applying the neutrino flavor polarization vector. Also, in a recent work, Ghai \textit{et al.} \cite{ghai2019neutrino} studied a coupling between ion-acoustic waves and a neutrino beam, considering flavor oscillations in dense relativistically degenerate plasmas, showing the significant modification of neutrino beam-driven instability growth due to flavor oscillations. Some further advancement of the neutrino beam-driven MHD theory, including the neutrino flavor oscillations, has been addressed in Ref. \cite{mendoncca2014influence,haas2013exact}. The effect of two-flavor oscillations was recently considered by Chatterjee \textit{et al.} \cite{chatterjee2023neutrino} to study oblique magnetosonic waves and resonant instabilities in neutrino beam-driven magnetoplasmas. The study revealed that two-flavor oscillations enhance instability growth rates of the oblique magnetosonic waves. However, the shear-Alfv{\'e}n waves remain unaffected by the neutrino beam and flavor oscillations. 
\par 
In a recent work, Misra \textit{et al.} \cite{misra2025} have shown that the Coriolis force of rotating fluids can couple oblique magnetosonic and Alfv{\'e}n waves and significantly influence the wave dispersion in classical magnetoplasmas. By considering an NMHD model with the effects of neutrino beam and two-flavor oscillations, we show that this force significantly alters neutrino-driven shear Alfv{\'e}n and oblique magnetosonic waves and instabilities by creating new couplings and wave modes, not reported before, that resonantly interact with the neutrino beam. These effects are pivotal in extreme astrophysical environments, such as those in the core collapse of supernovae and the dynamics of rotating neutron stars. The coupling mechanism can play a vital role in the energy transfer between neutrinos and plasmas, helping revive stalled supernova shocks by modifying MHD instabilities. One of the novelties of this study is that, in contrast to previous works [See, e.g., \cite{chatterjee2023neutrino,haas2016neutrino}], which reported the existence of decoupled Alfv{\'e}n and magnetosonic waves and that Alfv{\'e}n waves are neither influenced by the neutrino beam nor by the neutrino flavor oscillations, neutrinos can influence Alfv{\'e}n waves to undergo instabilities. Also, not only are both the Alfv{\'e}n and magnetosonic waves coupled and modified, but the instabilities associated with them are significantly altered by the coupling effects of one another. Another novelty is that the slow magnetosonic growth rate exhibits an inverted bell-shaped curve by the coupling effect, having maxima at parallel and antiparallel directions to the magnetic field, instead of the double-hump instability growth having maxima at other directions noted in the previous works of \cite{chatterjee2023neutrino,haas2016neutrino}. 
\par 
The possible key aspects could be:
\begin{itemize}
\item Faster explosion than typical times of supernovae explosions, since the combined influences of the neutrino beam and neutrino flavor oscillations on coupled MHD waves further enhance the growth rates of instability, especially in strong magnetic fields.
\item Enhancement of the efficiency of neutrino heating by expanding the heating region and allowing accreting matter to spend more time in the gain region. 
\item A more efficient energy transfer from neutrinos to plasmas compared to non-rotating plasmas, thereby increasing the likelihood of reviving the stalled shock, enabling the explosion to proceed outward faster.   
\end{itemize}   
\begin{figure}
    \includegraphics[width=3in,height=2.5in]{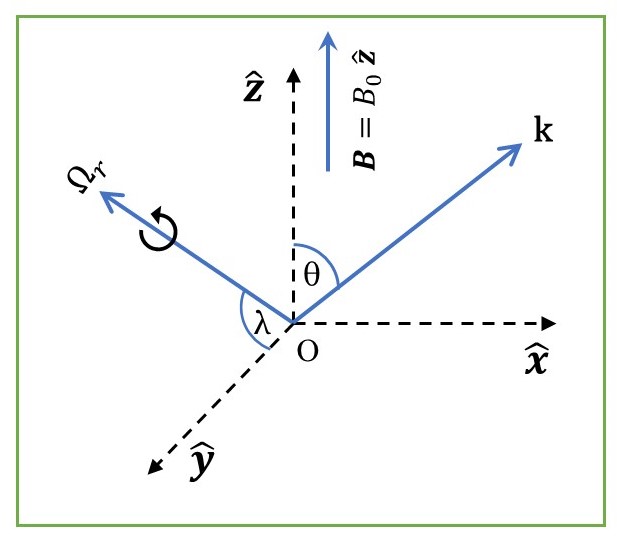}
	\caption{A schematic diagram showing the geometry of the magnetic field, rotational motion, and the wave vector. }
	\label{fig1-diag}
\end{figure}

\section{Physical model} \label{sec-bas-eq}
We consider the propagation of neutrino-driven MHD waves in homogeneous, fully ionized, highly conducting magnetized fluids, consisting of plasma electrons and ions, and assume that the fluids rotate about an axis in the $yz$ plane with a constant frequency $\Omega_r$ and interact weakly with electron and muon neutrinos via the electroweak force, $F_\nu$. We also assume that the plasma is immersed in a static magnetic field,  $\mathbf{B}=B_0 \hat{z}$, and the wave propagation vector is in the $xz$-plane, i.e., obliquely to the magnetic field. A geometry of the system configuration is depicted in Fig. \ref{fig1-diag}. In the single-fluid description of electrons and ions, the continuity and momentum balance equations read \cite{haas2016neutrino,chatterjee2023neutrino}
\begin{equation}
\label{eq-con}
    \frac{\partial \rho_{m}}{\partial t} + \mathbf{\nabla} \cdot (\rho_{m} \mathbf{U}) =0,
\end{equation}
\begin{equation}
\label{eq-mom}
    \frac{\partial \mathbf{U}}{\partial t} +\mathbf{U}\cdot \mathbf{\nabla U}=-\frac{\mathbf{\nabla} p}{\rho_{m}}  + \frac{(\mathbf{\nabla} \times \mathbf{B})\times \mathbf{B}}{\mu_{0}  \rho_{m}} +\frac{F_{\nu}}{m_{i}}+2 (\mathbf{U}\times \mathbf{\Omega}),
\end{equation}
where $\rho_{m}=m_{e} n_{e} +m_{i} n_{i} \approx nm_{i}$ (with $ n_{e}\approx n_{i}=n $) is the fluid mass density and $\mathbf{U}=(m_{e} n_{e} \mathbf{u}_{e} +m_{i} n_{i} \mathbf{u}_{i})/(m_{e} n_{e} +m_{i} n_{i}) \approx (m_{e} \mathbf{u}_{e} +m_{i} \mathbf{u}_{i})/m_{i} $ is the fluid velocity in the center-of-mass frame. Here, $m_{i(e)}$, $n_{i(e)}$, and $\mathbf{u}_{i(e)}$, respectively, denote the mass, number density, and velocity of ion (electron) fluids. Also, $\mu_{0}$ is the permeability of free space, $\mathbf{\Omega}$ is the angular velocity of the rotating fluid, $\mathbf{B}$ is the  magnetic field, and $p$ is total fluid pressure, given by, $\mathbf{\nabla}p=V_{s}^{2} \mathbf{\nabla} \rho_{m}$  with $V_{s}=\sqrt{k_B T_e/m_i}$ denoting the ion-acoustic speed (where $k_B$ is the Boltzmann constant and $T_e$ is the electron temperature). Furthermore, $\mathbf{F}_{\nu}= \sqrt{2} G_F (\mathbf{E}_{\nu}+ \mathbf{U} \times\mathbf{B}_{\nu} )$ is the neutrino-plasma weak interaction (electroweak) force with $G_F$ denoting the Fermi coupling constant and $E_{\nu}$ ($B_{\nu}$) the effective electric (magnetic field) induced by the weak interactions between neutrinos and  plasmas \cite{haas2016neutrino}, given by,
\begin{equation}
\label{eq-E-nu}
    E_{\nu}=- \mathbf{\nabla} N_e- \frac{1}{c^2} \frac{\partial}{\partial t} (N_e \mathbf{v}_e),
\end{equation}
\begin{equation}
    \mathbf{B}_{\nu}=\frac{1}{c^2} \mathbf{\nabla} \times( N_e \mathbf{v}_e).
\end{equation}
Here, $c$ is the speed of light in vacuum and $N_e$ ($\mathbf{v}_e$) is the number density (velocity) of electron neutrinos. 
\par 
The magnetic induction equation modified by the electroweak force $F_\nu$ is 
\begin{equation}
\label{eq-mag}
    \frac{\partial \mathbf{B}}{\partial t}=\mathbf{\nabla} \times \left( \mathbf{U} \times \mathbf{B} -\frac{\mathbf{F}_{\nu}}{e}\right),
\end{equation}
where $e$ is the elementary charge.
\par 
Next, we consider the coupling between plasma and two-flavor neutrino oscillations. For a coherent neutrino beam with an energy ${\cal E}_0$, the continuity equations for electron and muon neutrinos read \cite{mendoncca2014influence}
\begin{equation}
\label{eq-con-elc-nu}
    \frac{\partial N_e}{\partial t} +\mathbf{\nabla} \cdot (N_e \mathbf{v}_e)=\frac{1}{2} N \Omega_0 P_2,
\end{equation}
\begin{equation}
\label{eq-muon-nu}
 \frac{\partial N_{\mu}}{\partial t} +\mathbf{\nabla} \cdot (N_{\mu} \mathbf{v}_{\mu})=-\frac{1}{2} N \Omega_0 P_2, 
\end{equation}
where $N_{\mu}$ and $\mathbf{v}_{\mu}$ are, respectively, the number density and velocity of muon neutrinos. Also, $N=N_{e}+N_{\mu}$ is the total neutrino fluid density  and $P_2$ is a component of the flavor polarization vector $\mathbf{P}=(P_1, P_2, P_3)$ that corresponds to the neutrino coherence.
Furthermore, $\Omega_0=\omega_0 \sin (2 \theta_0)$, where $\omega_0=\delta m^2 c^4/2\hbar {\cal E}_0$ is the neutrino flavor oscillation frequency with $\delta m^2$ denoting the squared neutrino mass difference,  $\hbar$ the reduced Planck constant, and $\theta_0$ the neutrino oscillation mixing angle. The right-hand sides of Eqs. \eqref{eq-con-elc-nu} and \eqref{eq-muon-nu} show that the neutrino two-flavor oscillations contribute to the rates of electron and muon neutrino fluid densities. At the same time, the convective terms on the left-hand sides indicate the flows of electron and muon neutrino fluids into plasmas. 
\par 
From Eqs. \eqref{eq-con-elc-nu} and \eqref{eq-muon-nu}, integrating over the volume, we get  
\begin{equation} \label{eq-conser}
    \frac{d}{dt} \int \left( N_e+N_{\nu}\right) d^3 \mathbf{r}=-\int \mathbf{\nabla} \cdot (N_e \mathbf{v}_e+ N_{\nu} \mathbf{v}_{\nu}) d^3 \mathbf{r}.
\end{equation}
For the total neutrino fluid number densities to be conserved, the right side of Eq. \eqref{eq-conser}  must vanish. This, however, requires, e.g., decaying or periodic boundary conditions to apply.
\par The electron and muon neutrino momentum balance equations are \cite{mendoncca2014influence,chatterjee2023neutrino}  
\begin{equation}
\label{eq-Pe}
    \frac{\partial P_e}{\partial t} +\mathbf{v}_e \cdot \mathbf{\nabla} P_e=-\frac{\sqrt{2} G_F}{m_i} \mathbf{\nabla} \rho_m,
\end{equation}
\begin{equation}
 \frac{\partial P_{\mu}}{\partial t} +\mathbf{v}_{\mu} \cdot \mathbf{\nabla} P_{\mu}=0,
\end{equation}
where $P_{e(\mu)}={\cal E}_{e(\mu)} \mathbf{v}_{e(\mu)}/c^2$ is the electron (muon) neutrino relativistic momentum  and  ${\cal E}_{e(\mu)}=\left(P_{e(\mu)}^2c^2+m_{e(\mu)}^2 c^4\right)^{1/2}$  is the electron (muon) neutrino energy  with $m_{e(\mu)}$ denoting the electron (muon) neutrino mass.
\par 
Next, we require the time evolution equation for the flavor polarization vector $\mathbf{P}=(P_1, P_2, P_3)$ to complete the description for neutrino-plasma interactions. Thus, we have
\begin{equation}
\label{eq-pol-vec}
    \begin{split}
     &\frac{d P_{1}}{d t}=-\Omega_e P_2 , \\
     &\frac{d P_{2}}{d t}=-\Omega_e P_1-\Omega_0 P_3 ,\\
     &\frac{d P_{3}}{d t}=\Omega_0 P_2,
    \end{split}
\end{equation}
where the frequency, $\Omega_e=\omega_0[\cos(2\theta_0)-\sqrt{2} G_F n_e/(\hbar\omega_0)]$ is associated with the neutrino flavor oscillation frequency and the Fermi coupling.  Note that the total derivatives appearing in Eq. \eqref{eq-pol-vec} are, in general, different from each other. However, we assume that both the neutrino flavors have a constant streaming beam velocity ${\bf v}_0$, so that $d/dt\equiv \partial/\partial t+{\bf v}_0\cdot\nabla$. For time-dependent solutions of the polarization vector, the convective derivative can be disregarded in the linear analysis. An equilibrium state of Eq. \eqref{eq-pol-vec} can be defined as $P_{1}(0)=\Omega_0/\Omega_{\nu}$, $P_{2}(0)=0$, and $P_{3}(0)=\Omega_{e\mu}/\Omega_{\nu}=(N_{e0}-N_{\mu 0})/N_0$ such that $|P(0)|=1$. 
Equations \eqref{eq-pol-vec} are reducible to a single equation for $P_2$, i.e., 
\begin{equation}
\label{eq-2nd-order}
    \frac{d^2P_2}{dt^2}+\Omega_{\nu}^2 P_{2}=0,
\end{equation}
where $\Omega_{\nu}$ stands for the eigenfrequency of two-flavor oscillations, given by,
\begin{equation}
    \Omega_{\nu}^2=\Omega^2_{e0}+\Omega^2_{0}.
\end{equation}
An oscillating solution of Eq. \eqref{eq-2nd-order} can be presented in the form, $P_{2}(t)=A\sin\left( \Omega_{\nu}t\right)$ for some constant $A$.  
Equations~\eqref{eq-con}-\eqref{eq-pol-vec} constitute the basic equations for the excitation of neutrino-driven MHD waves in a homogeneous rotating magnetoplasma in the presence of two neutrino flavor oscillations.
It is important to note that the validity conditions for these simplified and ideal NMHD equations are almost the same as for high conductivity and nonrelativistic magnetoplasmas applicable for ideal MHD waves as described in the work of \cite{haas2016neutrino}. In addition, the NMHD model can well describe the core regions of core-collapse supernovae (CCSN), and PNSs provided $\lambda\ll L$, i.e., the deep regions within the PNSs where neutrinos scatter frequently to reach local thermodynamic and chemical equilibria with the surrounding plasma. Also, we will see in Sec. \ref{sec-Ins-Ana}, the model gives robust results for regions with the plasma number density $\sim10^{34}~\rm{m}^{-3}$, neutrino beam density  $\sim10^{37}~\rm{m}^{-3}$, and the magnetic field $\sim10^6-10^8$ T, or more.    
In Sec. \ref{Sec-Dispersion}, we will obtain a general dispersion relation for the Alfv{\'e}n and magnetosonic waves, and their coupling. 
\section{General dispersion relation} \label{Sec-Dispersion}
To obtain a general dispersion relation for coupled neutrino-driven MHD waves, we first split up the dependent variables into their equilibrium ($0$ value, or a quantity with suffix $0$) and perturbation parts (with suffix $1$) as follows:
\begin{equation}
\label{eq-per-exp}
\begin{split}
  &\rho_m=\rho_{m0}+\rho_{m1},~\mathbf{U}=0+\mathbf{U}_{1},~\mathbf{B}=\mathbf{B}_{0}+\mathbf{B}_{1},\\ 
  &N_{e}=N_{e0}+N_{e1},~N_{\mu}=N_{\mu0}+N_{\mu1},\\
  &\mathbf{v}_{e(\mu)}=\mathbf{v}_{e0(\mu_0)} (=\mathbf{v_0}) 
  +\mathbf{v}_{e1(\mu_1)}, 
\end{split}    
\end{equation}
where $N_0=N_{e0}+N_{\mu0}$ is the total equilibrium neutrino fluid density, and the magnitudes of the perturbed quantities are much smaller than
 the unperturbed parts. Next, we linearize Eqs. \eqref{eq-con}--\eqref{eq-pol-vec}, and assume the perturbed quantities to vary as plane waves with the wave frequency $\omega$ and the wave vector ${\bf k}$, i.e., in the form, $A(\mathbf{r},t)=A^\prime \exp\left(i{\bf k}\cdot\mathbf{r}-i\omega t\right)$, where $A^\prime$ is, in general, complex. Thus, from Eqs. \eqref{eq-con} and  \eqref{eq-mom}, we obtain the following linearized equations.  
\begin{equation}
\label{eq-lin-con}
    \omega \rho_{m1}=\rho_{m0}~ (\mathbf{k}\cdot \mathbf{U}_1),
\end{equation}
\begin{equation}
\label{eq-lin-mom}
    \omega \mathbf{U}_1=\frac{\mathbf{k} p_1}{\rho_{m0}}-\frac{(\mathbf{k}\times \mathbf{B}_1 )\times \mathbf{B}_0}{\mu_0 \rho_{m0} } +i\frac{\mathbf{F}_{\nu1}}{m_i}+i 2(\mathbf{U}_1\times \mathbf{\Omega}),
\end{equation}
with the linearized pressure, $p_1=V_s^2 \rho_1$  and the linearized form of the neutrino force, $\mathbf{F}_{\nu1}=\sqrt{2} G_F \mathbf{E}_{\nu1}$, noting that the term containing $\mathbf{B}_{\nu}$ is of the second order of smallness and thus neglected \cite{haas2017instabilities}.
\par 
The linearized form of the  magnetic induction equation [Eq. \eqref{eq-mag}] gives
\begin{equation}
 \omega \mathbf{B}_1+\mathbf{k}\times \left(\mathbf{U}_1 \times \mathbf{B}_0-\frac{\mathbf{F}_{\nu1}}{e}\right) =0.  
\end{equation}
From Eq. \eqref{eq-Pe}, one obtains 
\begin{equation}
\begin{split}
 (\omega-\mathbf{k}\cdot \mathbf{v}_0) P_{e1}&=\left[\mathbf{v}_{e1}+(1-\mathbf{v}_0^2/c^2)^{-1} \frac{\mathbf{v}_0.\mathbf{v}_{e1}}{c^2} \mathbf{v}_0\right]\\
 &=\sqrt{2} G_F \frac{\mathbf{k}}{m_i} \rho_{m1}.   
\end{split}    
\end{equation}
In the limit of non-relativistic fluid flow, $({v}_0\ll c)$, we obtain \cite{chatterjee2023neutrino}
\begin{equation}
\label{eq-ve1}
\begin{split}
    \mathbf{v}_{e1}=&\frac{\sqrt{2} G_F}{{\cal E}_0 (\omega-\mathbf{k}\cdot \mathbf{v}_0)}\\
    &\times \left[\frac{c^2 \mathbf{k} \rho_{m1}}{m_i}-\left(\frac{\mathbf{k}\cdot \mathbf{v}_0 \rho_{m1}}{m_i}-\frac{n_0 \omega \mathbf{v}_0\cdot \mathbf{U}_1}{c^2} \right)\right].
    \end{split}
\end{equation}
Substituting the value of $\rho_{m1}$ from Eq. \eqref{eq-lin-con} into Eq. \eqref{eq-ve1}, we have \cite{chatterjee2023neutrino}
\begin{equation}
 \mathbf{v}_{e1}=\frac{\sqrt{2} G_F}{{\cal E}_0(\omega-\mathbf{k}\cdot \mathbf{v}_0)}\frac{c^2 \rho_{m0} }{m_i \omega}
( \mathbf{k}\cdot \mathbf{U}_1) \mathbf{k}.
\end{equation}
Also, Eq. \eqref{eq-pol-vec} yields \cite{chatterjee2023neutrino}
\begin{equation}
    {P_2}_1=-i\frac{\sqrt{2} G_F \Omega_0 \omega}{(\omega^2-\Omega_{\nu}^2) m_i \hbar \Omega_{\nu}} \rho_{m1}.
\end{equation}
Now, plugging this expression for ${P_2}_1$ in Eq. \eqref{eq-con-elc-nu}, the linearized form of $N_{e1}$ is obtained as \cite{chatterjee2023neutrino}
\begin{equation}
\label{eq-Ne1}
\begin{split}
 N_{e1}=&N_{e0}  \frac{\sqrt{2} G_F }{{\cal E}_0(\omega-\mathbf{k}\cdot \mathbf{v}_0)^2}\frac{c^2 \rho_{m0} }{m_i \omega}
( \mathbf{k}\cdot \mathbf{U}_1) k^2 \\
&+\frac{\sqrt{2} G_F \Omega_0^2 N_0 \rho_{m0} }{2 m_i \hbar \Omega_{\nu}(\omega-\mathbf{k}\cdot \mathbf{v}_0)(\omega^2-\Omega_{\nu}^2)}
( \mathbf{k}\cdot \mathbf{U}_1).    
\end{split}
\end{equation}
The linearized form of the effective neutrino electric field, $\mathbf{E}_{\nu1}$ can be obtained from Eq.~\eqref{eq-E-nu} by substituting $N_{e1}$ from Eq.~\eqref{eq-Ne1} into it as
\begin{equation}
    \begin{split}
      F_{\nu1}=&\Bigg[V_N^2\frac{(\omega^2-c^2 k^2)}{(\omega-\mathbf{k}\cdot \mathbf{v}_0)^2}\\
      &+V_{\rm{osc}}^2\frac{\Omega_0^2\omega {\cal E}_0(\omega (\mathbf{k}\cdot \mathbf{v}_0)-c^2 k^2)}{2\hbar c^2 k^2\Omega_{\nu}(\omega-\mathbf{k}\cdot \mathbf{v}_0)(\omega^2-\Omega_{\nu}^2)}\Bigg]\frac{(\mathbf{k}\cdot \mathbf{U}_1)}{\omega}\mathbf{k},
    \end{split}
\end{equation}
where $V_N$ and $V_{\rm{osc}}$ are, respectively, the velocities of MHD perturbations associated with their interactions with the streaming of electron-neutrino beam and two neutrino flavor oscillations, given by, 
\begin{equation}
  V_N=\sqrt{\frac{2G_F^2 \rho_{m0} N_{e0}}{m_i^2 {\cal E}_0}},~ V_{\rm{osc}}=\sqrt{\frac{2G_F^2 \rho_{m0} N_{0}}{m_i^2 {\cal E}_0}}.   
\end{equation}
Finally, eliminating the nonzero perturbed quantities except $\mathbf{U}_1$, we obtain from Eq. \eqref{eq-lin-mom} the following general dispersion relation in vector form.
\begin{equation}
\label{eq-gen-dis-vec}
    \begin{split}
     &\omega \mathbf{U}_1=\left[V_s^2+V_A^2+V_N^2\frac{(c^2 k^2-\omega^2)}{(\omega-\mathbf{k}\cdot \mathbf{v}_0)^2} \right.\\
     &\left. +V_{\rm{osc}}^2\frac{\Omega_0^2\omega {\cal E}_0(c^2 k^2-\omega (\mathbf{k}\cdot \mathbf{v}_0))}{2\hbar c^2 k^2\Omega_{\nu}(\omega-\mathbf{k}\cdot \mathbf{v}_0)(\omega^2-\Omega_{\nu}^2)}\right]\frac{(\mathbf{k}\cdot \mathbf{U}_1)}{\omega}\mathbf{k}\\
     &+\frac{(\mathbf{k}\cdot \mathbf{V}_A)}{\omega}\bigg\{(\mathbf{V}_A \cdot \mathbf{k})\mathbf{U}_1-(\mathbf{V}_A \cdot \mathbf{U}_1)\mathbf{k}-(\mathbf{k}\cdot \mathbf{U}_1)\mathbf{V}_A\bigg\}\\
     &+i 2\Big(\mathbf{U}_1\times \mathbf{\Omega}\Big),   
    \end{split}
\end{equation}    
where $\mathbf{V}_A=\mathbf{B}_0/\sqrt{\mu_0 \rho_{m0}}$ is the Alfv{\'e}n velocity.  From Eq. \eqref{eq-gen-dis-vec}, we note that similar to Ref. \cite{chatterjee2023neutrino}, the resonances occur at frequencies, $\omega\approx {\bf k}\cdot {\bf v}_0$ due to the streaming neutrino beam and $\omega\approx \Omega_\nu$ due to coupling of MHD waves with neutrino two-flavor oscillations, also, due to the smallness of the Fermi constant $G_F$, the perturbations associated with $V_N^2$ and $V_{\rm{osc}}^2$ are assumed to be small. The MHD wave perturbations interacting resonantly with the neutrino beam and neutrino flavor oscillations will eventually lead to instabilities, which we will study in Sec. \ref{sec-Ins-Ana}.
Next, by redefining the sound speed, modified by the neutrino beam and two flavor oscillations, i.e.,
\begin{equation}
\label{eq-VS}
\begin{split}
  \widetilde{V}_s^2(\omega, k) =&V_s^2+V_N^2\frac{(c^2 k^2-\omega^2)}{(\omega-\mathbf{k}\cdot \mathbf{v}_0)^2}\\
   &+V_{\rm{osc}}^2\frac{\Omega_0^2\omega {\cal E}_0(c^2 k^2-\omega (\mathbf{k}\cdot \mathbf{v}_0))}{2\hbar c^2 k^2\Omega_{\nu}(\omega-\mathbf{k}\cdot \mathbf{v}_0)(\omega^2-\Omega_{\nu}^2)},  
\end{split}   
\end{equation}
Eq. \eqref{eq-gen-dis-vec} can be put in the following simplified form.
\begin{equation}
\label{eq-gen-dis-vec-sim}
    \begin{split}
     \omega \mathbf{U}_1=&\Big(\widetilde{V}_s^2+V_A^2\Big)\frac{(\mathbf{k}\cdot \mathbf{U}_1)}{\omega}\mathbf{k}+\frac{(\mathbf{k}\cdot \mathbf{V}_A)}{\omega}\Big\{(\mathbf{V}_A \cdot \mathbf{k})\\
     &\mathbf{U}_1-(\mathbf{V}_A \cdot \mathbf{U}_1)\mathbf{k}-(\mathbf{k}\cdot \mathbf{U}_1)\mathbf{V}_A\Big\}+i 2\big(\mathbf{U}_1\times \mathbf{\Omega}\big).   
    \end{split}
\end{equation}  
Equation \eqref{eq-gen-dis-vec-sim} is the vector form of the dispersion relation for neutrino-driven MHD waves, coupled to two-flavor oscillations, in rotating magnetoplasmas that interact with the streaming neutrino beam.
So far, we have considered wave propagation in an arbitrary direction relative to the static magnetic field. Next, without loss of any generality, we assume the wave propagation vector in the $xz$-plane and the angular velocity in the $yz$-plane (See Fig.~\ref{fig1-diag}), i.e.,  $\mathbf{k}=(k \sin\theta, 0, k \cos\theta)$ and $\mathbf{\Omega}=(0, \Omega_r \cos \lambda, \Omega_r \sin\lambda)$, where $\Omega_r$ is the rotational frequency. With these assumptions, Eq. \eqref{eq-gen-dis-vec-sim} gives three homogeneous linear equations for the perturbed velocity components $U_{x1}$, $U_{y1}$, and $U_{z1}$. Looking for the nonzero solutions of them, we obtain the following linear dispersion relation for neutrino-driven MHD waves in rotating magnetoplasmas.
\begin{equation}
\label{eq-dis-xzplane}
    \begin{split}
        &\left(\omega^2-k^2\cos^2\theta V_A^2\right) \Big[\omega^4-\Big(\Tilde{V}_s^2+V_A^2+\frac{4\Omega_r^2\cos^2 \lambda}{k^2}\Big)k^2\omega^2\\
        &+k^4\cos^2\theta V_A^2 \Tilde{V}_s^2\Big]-4\Omega_r^2\sin^2 \lambda \omega^2\left(\omega^2-k^2\cos^2\theta \Tilde{V}_s^2\right)=0.
    \end{split}
\end{equation}
\par 
 In the absence of the Coriolis force, or the rotational effect (the term proportional to $\Omega_r$), Eq. \eqref{eq-dis-xzplane} reduces to the same dispersion relation as in Ref. \cite{chatterjee2023neutrino}. Also, by disregarding the effects of neutrino flavor oscillations and the Coriolis force, one can recover the same dispersion relation as in Ref. \cite{haas2017instabilities}. Thus, Eq. \eqref{eq-dis-xzplane} generalizes and advances the previous works  \cite{haas2017instabilities,chatterjee2023neutrino}. 
From Eq. \eqref{eq-dis-xzplane}, we also note that while the first factor of the first term on the left side corresponds to shear Alfv{\'e}n waves, the second factor gives rise to slow and fast oblique magnetosonic modes. It is also interesting to observe that the shear Alfv{\'e}n and oblique magnetosonic waves become coupled and modified by the presence of the term proportional to $\Omega_r$ due to fluid rotation in the $yz$-plane, i.e., obliquely to the external magnetic field. In the absence of the fluid rotation or the Coriolis force, or if, in particular, $\lambda=0$, i.e., the fluid rotational axis is perpendicular to the magnetic field, the shear Alfv{\'e}n and oblique magnetosonic waves become decoupled. Their dispersion relations, respectively, are given by 
\begin{equation}
\omega^2=k^2\cos^2\theta V_A^2, \label{eq-alfven}
\end{equation}
\begin{equation}
\omega^4-\left(\widetilde{V}_s^2+V_A^2+\frac{4\Omega_r^2}{k^2}\right)k^2\omega^2+k^4\cos^2\theta V_A^2 \widetilde{V}_s^2=0, \label{eq-magnetos}
 \end{equation}
where the term proportional to $\Omega_r$ is nonzero if the influence of the Coriolis force is retained, otherwise zero. Equations \eqref{eq-alfven} and \eqref{eq-magnetos} agree with Ref. \cite{chatterjee2023neutrino} without the term proportional to $\Omega_r$.
 Thus, for couplings to occur, we must have $0<\lambda\leq \pi/2$. In this case, the coupling will result in modifications to both the shear Alfv{\'e}n and oblique magnetosonic waves that may lead to a phenomenon of wave mixing where they begin to exhibit characteristics of both. In the nonlinear regime, this coupling may lead to energy propagation across magnetic field lines and its redistribution in different directions. Such processes are crucial for wave heating and energy transport in astrophysical plasmas. Furthermore, the coupling may lead to new observed characteristics, including the generation of fast magnetosonic waves from initial Alfv{\'e}n waves. 
 Before proceeding to explore the instability characteristics of the neutrino-driven MHD waves in the presence of the Coriolis force, which is the prime interest of the current investigation, we first discuss two particular cases of interest, namely when $\mathbf{k \perp \mathbf{B}_0}$ for $\theta=\pi/2$  and $\mathbf{k \parallel \mathbf{B}_0}$ for $\theta=0$.
\par 
  For wave propagation perpendicular to the magnetic field, the dispersion relation \eqref{eq-dis-xzplane} gives the following fast magnetosonic wave as the only propagating mode.
 \begin{equation}
\omega^2-\Big(\Tilde{V}_s^2+V_A^2\Big)k^2-4\Omega_r^2=0.    
 \end{equation}
We note that the fast magnetosonic mode becomes dispersive by the influence of the Coriolis force due to rotation of fluids with the phase velocity, $\omega/k=\left[\left(\Tilde{V}_s^2+V_A^2\right)+{4\Omega_r^2}/{k^2}\right]^{1/2}$ being modified by the term proportional to $\Omega_r$. It follows that both the wave frequency and the phase velocity of fast magnetosonic waves are enhanced relative to those in nonrotating fluids. While the wave frequency increases with $k$, the phase velocity achieves a maximum value in the long-wavelength limit ($k\to0$) but then decreases with increasing values of $k~(>0)$. 
\par 
On the other hand, for wave propagation parallel to the magnetic field $\mathbf{B}_0$, i.e., $\theta=0$, Eq. \eqref{eq-dis-xzplane} gives the following coupled modes with mixed characteristics of both the Alfv{\'e}n and magnetosonic waves.
 \begin{equation}
 \begin{split}
  &\left(\omega^2-k^2 V_A^2\right) \bigg[\omega^4-\Big(\Tilde{V}_s^2+V_A^2+\frac{4\Omega_r^2}{k^2}\Big)k^2\omega^2+k^4 V_A^2 \Tilde{V}_s^2\bigg]\\
  &-4\Omega_r^2\sin^2 \lambda\left(\omega^2-k^2 \Tilde{V}_s^2\right) \omega^2=0.      
 \end{split} 
 \end{equation}
As mentioned before, such a coupling occurs for the rotational angle $\lambda$ satisfying $0<\lambda\leq 90\degree$. We limit here the detailed discussion about the characteristics of these particular modes rather  we focus on the MHD wave instability in a more general situation in Sec. \ref{sec-Ins-Ana}.

\section{Instability Analysis} \label{sec-Ins-Ana}
In Sec. \ref{sec-bas-eq}, we observed that the shear Alfv{\'e}n and oblique magnetosonic waves (fast and slow) become coupled due to the Coriolis force, associated with the rotational frequency $\Omega_r$ and obliqueness angle $\lambda$. This coupling significantly alters these waves, changing them from purely shear Alfv{\'e}nic to mixed magneto-acoustic-Alfv{\'e}nic types. As a result, the simple, non-dispersive nature of Alfv{\'e}n waves is also affected. In the nonlinear regime, such coupling may cause energy transfer between Alfv{\'e}n and magnetosonic waves, but this is beyond our present study, in Secs. \ref{sec-magn} and \ref{sec-alfv}, we will study the instabilities associated with these modified magnetosonic and Alfv{\'e}n waves separately. 
\subsection{Instability growth rates for magnetosonic waves} \label{sec-magn}
To obtain the instability growth rate for the neutrino-driven magnetosonic waves, modified by the effects of coupling with Alfv{\'e}n wave modes, we rewrite the dispersion relation in Eq.~\eqref{eq-dis-xzplane} by diving the Alfv{\'e}nic mode factor as
\begin{equation}
\label{eq-magtesonic}
    \begin{split}
     &\omega^4-\left(V_s^2+V_A^2+\frac{4\Omega_r^2\cos^2\lambda}{k^2}\right)k^2\omega^2+k^4\cos^2\theta V_A^2V_s^2 \\
     &=4\Omega_r^2\sin^2\lambda \frac{\left(\omega^2-k^2\cos^2\theta V_s^2\right)\omega^2}{\left(\omega^2-k^2\cos^2\theta V_A^2\right)}\\
     &+ \left[\omega^2k^2-k^4\cos^2\theta V_A^2-\frac{4\Omega_r^2\sin^2 \lambda \cos^2\theta k^2\omega^2}{\left(\omega^2-k^2\cos^2\theta V_A^2\right)}\right]\\
      &\times\left[V_N^2\frac{(c^2 k^2-\omega^2)}{(\omega-\mathbf{k}\cdot \mathbf{v}_0)^2}\right.\\ 
   &\left.  +V_{\rm{osc}}^2\frac{\Omega_0^2\omega {\cal E}_0(c^2 k^2-\omega (\mathbf{k}\cdot \mathbf{v}_0))}{2\hbar c^2 k^2\Omega_{\nu}(\omega-\mathbf{k}\cdot \mathbf{v}_0)(\omega^2-\Omega_{\nu}^2)}\right].
    \end{split}
\end{equation}
From Eq. \eqref{eq-magtesonic}, we note that not only the Alfv{\'e}nic factor contributes to the coupling terms proportional to $\Omega_r$, it also modifies the terms associated with the neutrino beam (the term proportional to $V_N^2$) and neutrino two-flavor oscillations (the term proportional to $V_{\rm{osc}}^2$). We assume that the Alfv{\'e}nic contribution to the magnetosonic wave dispersion is small. Also, since the perturbations associated with the neutrino contributions are small and they resonantly interact with the neutrino beam and flavor oscillations, we consider 
\begin{equation}
    \omega=\widetilde{\Omega}_m+\delta\omega,
\end{equation}
where $\delta\omega$ is a small part of the wave frequency that appears due to the neutrino contributions to the wave dispersion such that  $|\delta\omega|\ll \widetilde{\Omega}_{m}$,
and the following double resonance conditions:
\begin{equation}
\label{eq-dou-res}
    \omega=\Omega_{\nu} \approx \widetilde{\Omega}_m=\mathbf{k}\cdot \mathbf{v}_0.
\end{equation}
Here, $\widetilde{\Omega}_m$ is a solution of the following dispersion equation (without the coupling and neutrino effects). 
\begin{equation} \label{eq-mag-FS}
 \omega^4-\Big(V_s^2+V_A^2+\frac{4\Omega_r^2\cos^2\lambda}{k^2}\Big)k^2\omega^2+k^4\cos^2\theta V_A^2V_s^2=0.
\end{equation}
From Eq. \eqref{eq-mag-FS}, the frequencies of fast and slow magnetosonic modes can be obtained by assuming
\begin{equation}
    \omega=\widetilde{\Omega}_{m\pm}=k V_{\pm},
\end{equation}
where the `plus' and `minus' sign, respectively, correspond to the fast and slow magnetosonic modes. The corresponding phase velocities are given by
\begin{equation}
\begin{split}
 &V_{\pm}=\Bigg[\frac{1}{2}\Bigg\{\Big(V_s^2+V_A^2+\frac{4\Omega_r^2\cos^2\lambda}{k^2}\Big)\pm V_{\rm{AS}}\Bigg\}\Bigg]^{1/2} ,  
 \end{split}
\end{equation}
where $V_{\rm{AS}}$, accounting for the acoustic, Alv{\'e}nic, and Coriolis force effects, is given by 
\begin{equation}
    V_{\rm{AS}}=\sqrt{\Big(V_s^2+V_A^2+\frac{4\Omega_r^2\cos^2\lambda}{k^2}\Big)^2-4\cos^2\theta V_A^2 V_s^2}.
\end{equation}
We note that in contrast to non-rotating plasmas \cite{chatterjee2023neutrino}, both the fast and slow magnetosonic waves become dispersive by the influence of the Coriolis force that introduces a frequency-dependent correction $(\propto \Omega_r^2)$ to the wave dynamics.
Next, applying the double resonance condition [See Eq. \eqref{eq-dou-res}] together with the nonrelativistic fluid flow condition, $V^2_{\pm}\ll c^2$, we obtain from Eq. \eqref{eq-magtesonic} the following expression for the frequency correction $\delta\omega$.
\begin{equation}
\label{eq-mag-delomega1}
    \begin{split}
    \delta\omega&= 4\Omega_r^2\sin^2\lambda \frac{\Big(V_{\pm}^2-\cos^2\theta V_s^2\Big)V_{\pm}^2}{2V_{\pm} V_{\rm{AS}} \Big(V_{\pm}^2-\cos^2\theta V_A^2\Big)k }\\
    &+ \Big\{V_{\pm}^2-\cos^2\theta V_A^2-\frac{4\Omega_r^2\sin^2 \lambda \cos^2\theta V_{\pm}^2}{\Big(V_{\pm}^2-\cos^2\theta V_A^2\Big)k^2}\Big\} \times\\
    &\Big\{\frac{V_N^2 c^2 k^3}{2V_{\pm} V_{\rm{AS}}}+\frac{G_F^2\rho_{m0} N_0\Omega_0^2}{4m_i^2\hbar V_{\pm}^2 V_{\rm{AS}}}\Big\}\frac{1}{(\delta\omega)^2}. 
    \end{split}
\end{equation}
Equation \eqref{eq-mag-delomega1} can be written in the following form.
	\begin{equation}
		\label{eq-mag-delomega2}
		\delta\omega= \bigg[l_{1}^{\pm} \Big(\frac{ \gamma_{\nu}^{\pm}+\gamma_{\rm{os}}^{\pm}}{\delta\omega^2}\Big)+ l_{2}^{\pm}\bigg],
	\end{equation}
	where
	\begin{equation}
	l_{1}^{\pm}= \left[V_{\pm}^2-\cos^2\theta V_A^2-\frac{4\Omega_r^2\sin^2 \lambda \cos^2\theta V_{\pm}^2}{\left(V_{\pm}^2-\cos^2\theta V_A^2\right)k^2}\right],   
	\end{equation}
	\begin{equation}
		l_{2}^{\pm}= 4\Omega_r^2\sin^2\lambda \frac{\Big(V_{\pm}^2-\cos^2\theta V_s^2\Big)V_{\pm}^2}{2V_{\pm} V_{\rm{AS}} \Big(V_{\pm}^2-\cos^2\theta V_A^2\Big)k }. 
	\end{equation}
	The quantities $\gamma_{\nu}^{\pm}$ and $\gamma_{\rm{os}}^{\pm}$ in Eq. \eqref{eq-mag-delomega2} appear due to interactions of MHD waves with the streaming neutrino beam and the coupling of MHD waves with neutrino two-flavor oscillations, respectively, given by,
	\begin{equation}
		\gamma_{\nu}^{\pm}=\frac{V_N^2 c^2 k^3}{2V_{\pm} V_{\rm{AS}}}, 
			\gamma_{\rm{os}}^{\pm}= \frac{G_F^2\rho_{m0} N_0\Omega_0^2}{4m_i^2\hbar V_{\pm}^2 V_{\rm{AS}}}.  
\end{equation}
\par 
Next, in the weak coupling limit, $|\Omega_r \sin\lambda| \ll 1$, Eq. \eqref{eq-mag-delomega1} gives
\begin{equation}
    \delta\omega=\Big(V_{\pm}^2-\cos^2\theta V_A^2\Big)^{1/3}\Bigg[\frac{V_N^2 c^2 k^3}{2V_{\pm} V_{\rm{AS}}}
    +\frac{G_F^2\rho_{m0} N_0\Omega_0^2}{4m_i^2\hbar V_{\pm}^2 V_{\rm{AS}}}\Bigg]^{1/3}, 
\end{equation}
or, we write
\begin{equation}
		\begin{split}
			\delta\omega={l_{3}^{\pm}}^{1/3}\Big[ \gamma_{\nu}^{\pm}+\gamma_{\rm{os}}^{\pm}\Big]^{1/3},
		\end{split}   
	\end{equation}
	where $l_{3}^{\pm}=(V_{\pm}^2-\cos^2\theta V_A^2)$. 
Plugging this expression for $\delta\omega$ into the right side of Eq. \eqref{eq-mag-delomega2}, we obtain
	\begin{equation}
		\begin{split}
			\delta\omega=\Bigg[l_{1}^{\pm} \Big( \gamma_{\nu}^{\pm}+\gamma_{\rm os}^{\pm}\Big)+ l_{2}^{\pm} {l_{3}^{\pm}}^{2/3}\Big( \gamma_{\nu}^{\pm}+\gamma_{\rm{os}}^{\pm}\Big)^{2/3}\Bigg]^{1/3}.
		\end{split}
\end{equation}
\par 
Finally, the growth rates of instability [$\gamma=\Im(\delta\omega)$] for fast and slow magnetosonic modes  can be obtained as 
	\begin{equation}\label{eq-gam-mag}
		\begin{split}
			\gamma_{\pm}=\frac{\sqrt{3}}{2}\Bigg[& l_{1}^{\pm} \Big( \gamma_{\nu}^{\pm}+\gamma_{\rm os}^{\pm}\Big)+ l_{2}^{\pm} {l_{3}^{\pm}}^{2/3}\Big( \gamma_{\nu}^{\pm}+\gamma_{\rm os}^{\pm}\Big)^{2/3}\Bigg]^{1/3}.
		\end{split}
\end{equation} 
From Eq. \eqref{eq-gam-mag}, it is evident that compared to Ref. \cite{chatterjee2023neutrino}, the expression for the growth rates of the magnetosonic modes become significantly modified by the influence of the Coriolis force (terms proportional to $\Omega_r$) and contribution from the Alfv{\'e}n mode due to the coupling by this force. 
Next, we numerically study Eq. \eqref{eq-gam-mag} for the profiles of the growth rates of instabilities of both the fast and slow magnetosonic modes influenced by the coupling effect of the Alfv{\'e}n mode due to the Coriolis force in rotating magnetoplasmas. We consider typical plasma parameters that are characteristic of the type II CCSN SN1987A \cite{haas2017instabilities}. In these environments, the expected fluid flow comprises $10^{58}$ neutrinos of all flavors with the streaming energy, ${\cal E}_0 \sim 10-15$ MeV. Also, the neutrino beam density is $N_0\sim 10^{34}-10^{37}$ m$^{-3}$ with a strong magnetic field in the range, $B_0 \sim 10^6-10^8$ T. However, in the present analysis, we have chosen 
 $n_0\sim10^{34}$m$^{-3}$, $N_{e0}\sim10^{37}$m$^{-3}$, $N_0\sim5\times10^{37}$m$^{-3}$, and $T_e\sim0.1$ MeV. Furthermore, $\Delta m^2 c^4\sim3 \times 10^{-6}$ (eV)$^2$, $\sin2\theta\sim10^{-1}$, ${\cal E}_0\sim10$ MeV, $G_F=1.45\times 10^{-62}$ Jm$^{-3}$, $\Omega_r\sim5\times 10^{3}$ s$^{-1}$, and $B_0\sim10^{6}-10^{7}$ T.  
\begin{figure*}
	\centering
	\includegraphics[width=\textwidth]{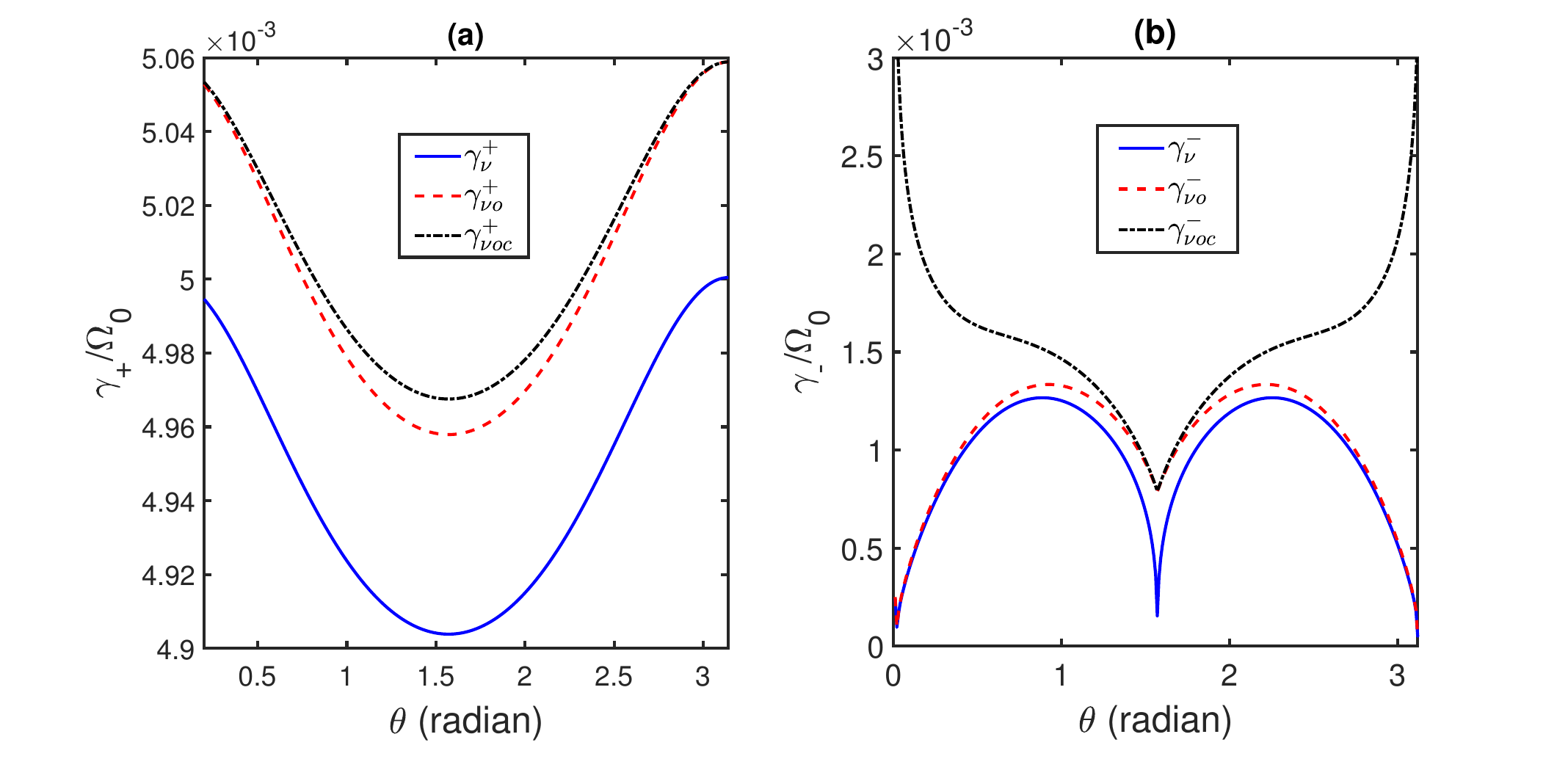}
         \includegraphics[width=\textwidth]{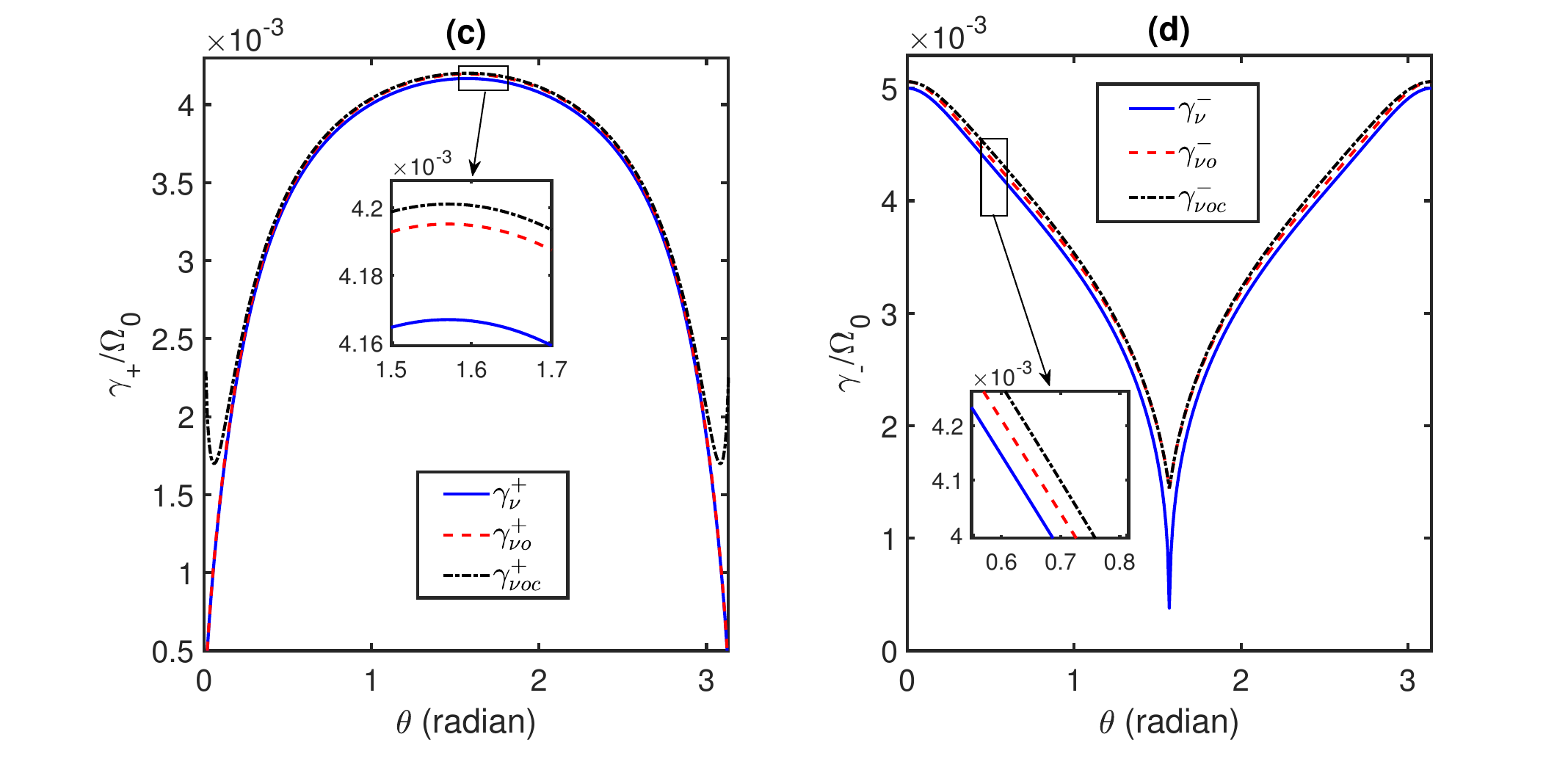}
	\caption{Instability growth rates corresponding to fast [with a plus sign; subplots (a) and (c)]  and slow [with a minus sign; subplots (b) and (d)] magnetosonic waves are shown against the propagation angle $\theta$.  The blue solid ($\gamma_\nu^{\pm}$), red dashed ($\gamma_{\nu o}^{\pm}$) and black dashed-dotted ($\gamma_{\nu oc}^{\pm}$) lines correspond to the instability growth rates when (i) only the neutrino beam effects are present, (ii) only the neutrino beam and two-flavor effects are present, and (iii) the neutrino beam, two-flavor effects, and the effects of coupling with the Alfv{\'e}n modes in the presence of the Coriolis force are present, respectively.  The magnetic field strength for subplots (a) and (b) [subplots (c) and (d)] is $B_0=5\times 10^{6}$ T ($B_0=2\times 10^{7}$ T). The other fixed parameter values are $n_0=10^{34}$ m$^{-3}$, $k=10^2$ m$^{-1}$, $\lambda=\pi/4$, and $\Omega_r=5\times 10^{3}$ s$^{-1}$.}
	\label{fig2-fsmag_b}
\end{figure*}

\begin{figure*}
	\centering
	\includegraphics[width=\textwidth]{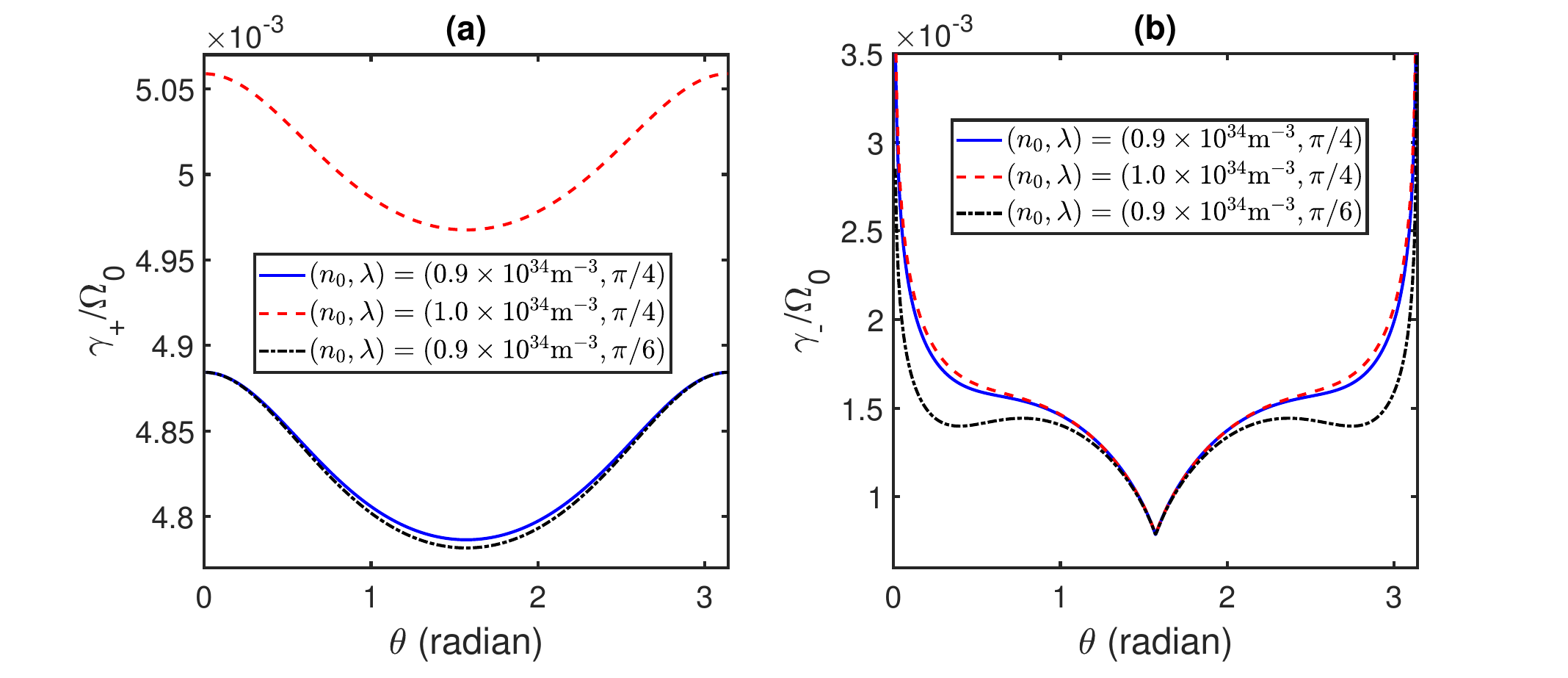}
	\caption{Instability growth rates [when all the effects as for $\gamma_{\rm{\nu oc}}^{\pm}$ in Fig. \ref{fig2-fsmag_b} are included] of fast [subplot (a)] and slow [subplot (b)] oblique magnetosonic waves are shown against the propagation angle $\theta$ for different values of the plasma density $n_0$ and the obliqueness of the axis of rotation $\lambda$ as in the legends.  The magnetic field is $B_0=5\times 10^{6}$ T. All other fixed parameter values are as for Fig. \ref{fig2-fsmag_b}.}
	\label{fig3-fsmag_n}
\end{figure*}
\par 
 The key parameters are $\theta$ (the obliqueness of the wave propagation direction with the magnetic field), $B_0$ (the external magnetic field), $n_0$ (the plasma number density), and $\lambda$ (the fluid rotational angle). To exhibit the profiles of the growth rates and compare the results with those in Ref. \cite{chatterjee2023neutrino}, we plot $\gamma_{\pm}$ against $\theta$. We use variations of the magnetic field similar to Ref. \cite{chatterjee2023neutrino}, and exhibit the results in Fig.~\ref{fig2-fsmag_b}. Subplots [(a) and (b)] and subplots [(c) and (d)], respectively, show the growth rates for the fast and slow magnetosonic modes with $B_0=5 \times 10^{6}$ T and $B_0=2 \times 10^{7}$ T.
The subplots contain three different instability growth curves. The (i) solid (blue), (ii) dashed (red), and (iii) dash-dotted (black) curves, respectively, correspond to the following: (i) only the impact of the neutrino beam is present (without the Coriolis force); (ii) the influences of both the neutrino beam and two-flavor oscillations are present (without the Coriolis force); and (iii) the influences of both the neutrino beam and two-flavor oscillations are present together with the coupling effects of the shear Alfv{\'e}n mode due to the Coriolis force. The solid and dashed curves match those in Ref. \cite{chatterjee2023neutrino}. The dash-dotted curves exhibit increased growth rates compared to the solid and dashed curves due to the influences of the Coriolis force and wave coupling. The qualitative features of the growth rates for the fast and slow magnetosonic modes, with two different magnetic field strengths, remain the same as in Ref. \cite{chatterjee2023neutrino}. There is one exception: the particular case of $\gamma_{-}$ with $B_0=5 \times 10^{6}$ T corresponding to (iii) above [Subplot (b), dash-dotted curve].  
\par 
To explain the features of the growth rates in more detail, from subplot (a) of Fig. \ref{fig2-fsmag_b}, when the magnetic field strength is relatively low, i.e., $B_0=5 \times 10^{6}$ T, inverted bell-shaped instability growth curves are obtained similar to Ref. \cite{chatterjee2023neutrino}. The growth rate decreases with the obliqueness angle of propagation $(\theta)$ in the domain $0<\theta< \pi/2$, attains a minimum value at $\theta=\pi/2$, and thereafter increases with $\theta$ in the domain $\pi/2<\theta< \pi$. Thus, we conclude that the fast magnetosonic mode becomes more unstable in the parallel ($\theta=0$) and antiparallel ($\theta=\pi$) directions of wave propagation than in the perpendicular ($\theta=\pi/2$) direction. We also observe that the growth rate gets significantly enhanced by the combined influences of the neutrino beam and two-flavor oscillations (Compare the solid and dashed curves). By preserving the same qualitative features, the growth rate is further enhanced by the coupling effects of the Alfv{\'e}n mode in the presence of the Coriolis force in addition to the effects of the neutrino beam and two-flavor oscillations (See the dash-dotted curve).  
 Next, for slow magnetosonic waves, Fig. \ref{fig2-fsmag_b} (b) shows a symmetric double-hump instability growth. The growth curves correspond to the same magnetic field strength as in Fig. \ref{fig2-fsmag_b} (a). Such a behavior occurs when the impacts of only the neutrino beam (solid curve), or both the neutrino beam and two-flavor oscillations (dashed curve), are present. We obtain the maximum growth rates at $\theta=\pi/4$ and $3\pi/4$, with nonzero minima at $\theta=0$ and $\theta=\pi$. We also note that $\gamma_{-}$ is not defined at $\theta=\pi/2$, where $\Omega_\nu=0$. However, close to this angle, $\gamma_{-}$ attains another nonzero minimum. Thus, near the propagation angles $\theta=0,~\pi/2,$ and $\pi$, the slow magnetosonic mode tends to become stable. A novel feature of the growth rate $\gamma_{-}$ is observed due to the influences of coupling with the Alfv{\'e}n mode from the Coriolis force, in addition to the effects of the neutrino beam and two-flavor oscillations (see the dash-dotted curve).
\par 
On the other hand, when the magnetic field is relatively strong compared to the Figs. \ref{fig2-fsmag_b} (a) and (b), i.e., $B_0=2 \times 10^{7}$ T, we see a behavioral change in the instability curves [See subplots (c), (d) of Fig. \ref{fig2-fsmag_b}]. Similar to Ref. \cite{chatterjee2023neutrino}, the bell-shaped growth curve for the fast magnetosonic mode has minima near $\theta=0$ and $\theta=\pi$ and a maximum near $\theta=\pi/2$ [See subplot (c)]. However, for the slow magnetosonic mode, the growth rate appears to be in the $\gamma$-shape, having maxima near $\theta=0$ and $\theta=\pi$ and a minimum near $\theta=\pi/2$ (Note that $\gamma_{-}$ is undefined at $\theta=\pi/2$ where $\Omega_\nu=0$) [See subplot (d)]. In both the cases of subplots (c) and (d), the growth rates increase as we successively enhance the effects from the neutrino beam to neutrino beam and flavor oscillations to neutrino beam, flavor oscillations, and the Alfv{\'e}n mode-coupling by the Coriolis force.    
\par 
Given a magnetic field strength at $B_0=5 \times 10^{6}$ T, we also examine the influences of the plasma number density $(n_0)$ and the rotational angle $(\lambda)$ on the profiles of the growth rates for the neutrino-driven fast [Subplot (a)] and slow [Subplot (b)] magnetosonic modes that become modified by the Alfv{\'e}n mode-coupling due to the Coriolis force. We display the results in Fig.~\ref{fig3-fsmag_n}.
We observe that while the slow mode shows a small increase in growth rate with a slight increase in plasma density, the growth rate of the fast mode becomes significantly enhanced by this density change (See the solid and dashed curves). On the other hand, a small decrease in the rotational angle $\lambda$ gives rise to a small dip in the growth curve for the fast mode, whereas a significant decrease occurs in the growth rate for the slow mode over the entire domain of $\theta$ except near $\theta=\pi/2$. 
 \par 
   From the instability analysis of the neutrino-driven modified magnetosonic waves, we conclude that the coupling effect by the shear Alfv{\'e}n wave due to the Coriolis force is ultimately responsible for enhancing the instability growth rate, leading to faster MHD wave instability compared to that reported in Ref. \cite{chatterjee2023neutrino}. A novel feature is that, in contrast to Ref. \cite{chatterjee2023neutrino} or in the absence of the coupling effect, the growth rate for the slow magnetosonic mode can no longer attain minimum values near $\theta=0$ and $\theta=\pi$, but maxima there, indicating a stronger instability. The time scales of these instabilities are also shorter than the characteristic time scales of supernova explosions.  
\subsection{Instability growth rates for shear Alfv{\'e}n modes} \label{sec-alfv}
The neutrino-driven shear Alfv{\'e}n mode is an MHD mode altered by magnetosonic wave coupling via the Coriolis force. Without this coupling, the shear Alfv{\'e}n mode is stable and unaffected by neutrino effects \cite{chatterjee2023neutrino}. While previous investigations have not examined instabilities in these waves, this section presents novel results. To derive the growth rate, we rewrite the dispersion relation \eqref{eq-dis-xzplane} as follows:
\begin{equation}
\label{eq-alf}
   \begin{split}
   &\omega^2= V_A^2 k^2\cos^2\theta\\
   &+\frac{4\Omega_r^2\sin^2 \lambda \omega^2\left(\omega^2-k^2\cos^2\theta \Tilde{V}_s^2\right)}{\Big[\omega^4-\Big(\Tilde{V}_s^2+V_A^2+\frac{4\Omega_r^2\cos^2 \lambda}{k^2}\Big)k^2\omega^2+k^4\cos^2\theta V_A^2 \Tilde{V}_s^2\Big]},
    \end{split}  
\end{equation}
where the last term is the correction term that appears due to the Coriolis force effect for rotating fluids. Interestingly, because of this force, the correction term also incorporates the effects of the neutrino beam and neutrino two-flavor oscillations. Previous investigations (See, e.g., Refs. \cite{chatterjee2023neutrino,haas2013exact,haas2016neutrino,haas2017instabilities}) did not report such neutrino contributions to the shear Alfv{\'e}n wave. 
\par 
Next, similar to Sec. \ref{sec-magn}, we consider
\begin{equation}
\label{eq-alf-pertur}
    \omega=\widetilde{\Omega}_A+\delta\omega,
\end{equation}
where $\delta\omega$ is associated with the correction term  proportional to $\Omega_r$ in Eq. \eqref{eq-alf} such that $|\delta\omega|\ll \widetilde{\Omega}_{A}$, and $\widetilde{\Omega}_{A}$ is a solution of the following dispersion relation, which is obtained from Eq. \eqref{eq-alf} by applying the weak coupling condition $|\Omega_r \sin\lambda| \ll 1$, given by, 
\begin{equation}
 \omega=\widetilde{\Omega}_{A\pm}=\pm k\cos \theta V_A.
\end{equation}
Here, $\widetilde{\Omega}_{A\pm}$ is the frequency of the fast/slow shear Alfv{\'e}n mode. We also use the following double resonance condition for both the fast and slow modes.
\begin{equation}\label{eq-alf-res}
\omega=\Omega_\nu\approx \widetilde{\Omega}_A={\bf k}\cdot {\bf v}_0.
\end{equation}
\begin{figure*}
	\centering
	\includegraphics[width=\textwidth]{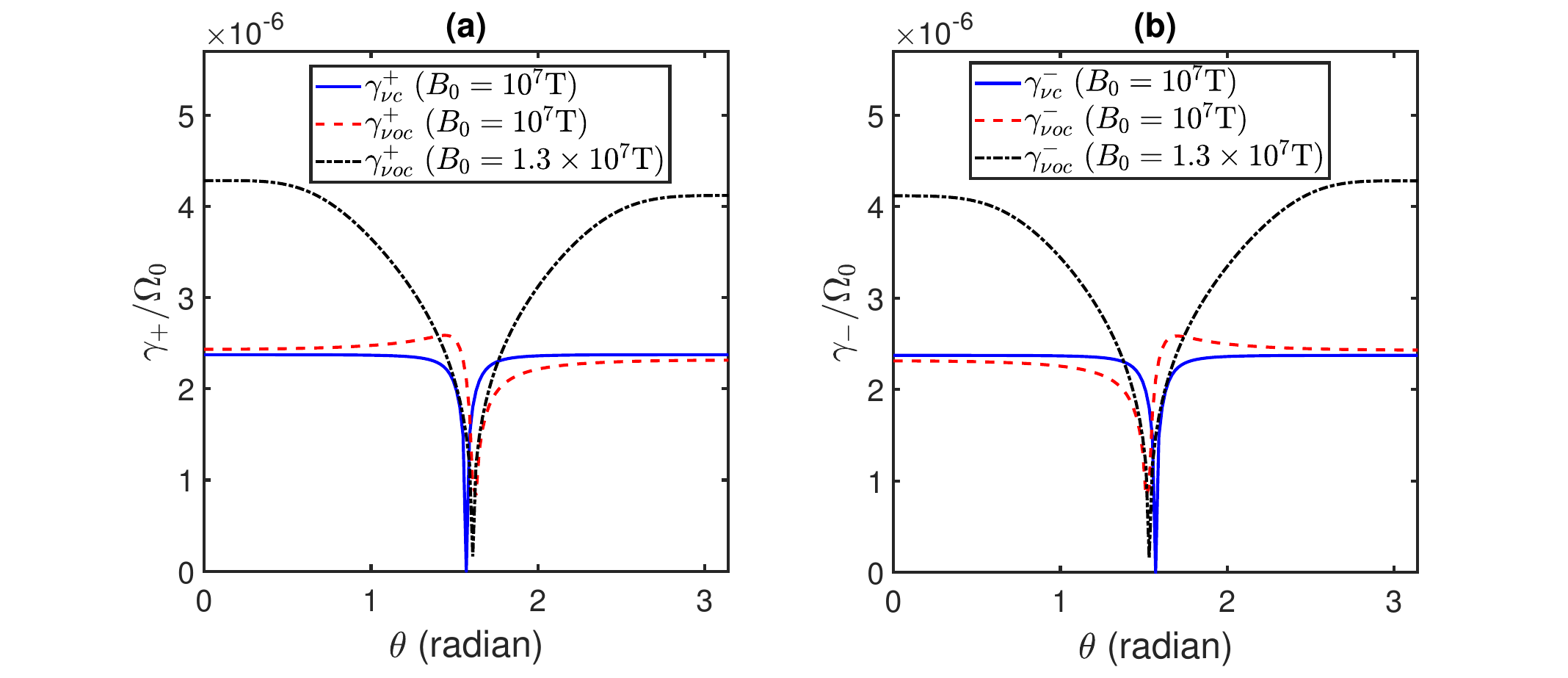}
	\caption{ Instability growth rates corresponding to fast [with a plus sign; subplot (a)]  and slow [with a minus sign; subplots (b)] shear Alfv{\'e}n modes are shown against the propagation angle $\theta$. The blue solid ($\gamma_{\nu c}^{\pm}$) and red dashed ($\gamma_{\nu oc}^{\pm}$) lines correspond to the instability growth rates when (i) only the neutrino beam effects and the effects of coupling with the magnetosonic modes in the presence of the Coriolis force are present and (ii) the neutrino beam, two-flavor effects, and the effects of coupling with the magnetosonic modes in the presence of the Coriolis force are present, respectively.    The black dash-dotted line is for the growth rate $\gamma_{\nu oc}^{\pm}$ with an increased magnetic field as in the legends. We have considered the rotational frequency as $\Omega_r=5\times 10^{3}$ s$^{-1}$ and other fixed parameter values as for Fig. \ref{fig3-fsmag_n}.}
	\label{fig4-gr-alfven}
\end{figure*}
\begin{figure*}
	\centering
	\includegraphics[width=\textwidth]{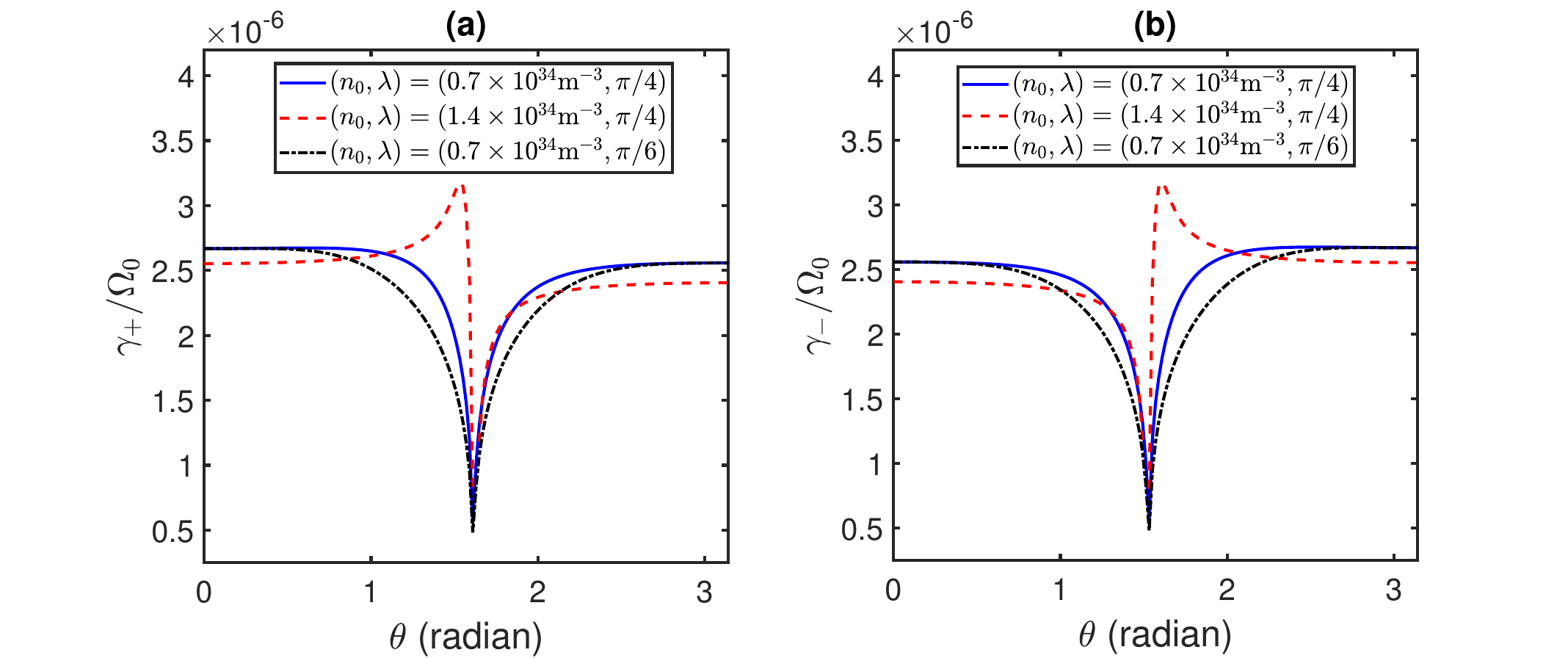}
	\caption{Instability growth rates [when all the effects as for $\gamma_{\rm{\nu oc}}^{\pm}$ in Fig. \ref{fig4-gr-alfven} are included] corresponding to fast [with a plus sign; subplot (a)]  and slow [with a minus sign; subplots (b)] shear Alfv{\'e}n modes are shown against the propagation angle $\theta$ for different values of the plasma density $n_0$ and the obliqueness of the angle of rotation $\lambda$ as in the legends. We have considered the magnetic field and the rotational frequency as  $B_0=10^7$ T and $\Omega_r=5\times 10^{3}$ s$^{-1}$. All other fixed parameter values are as for Fig. \ref{fig3-fsmag_n}.}
	\label{fig5-gr-alfven-n0}
\end{figure*}
\par 
Proceeding in the same way as in Sec. \ref{sec-magn}, we obtain from Eqs. \eqref{eq-alf}-\eqref{eq-alf-res} by applying the non-relativistic fluid flow condition $V_{A\pm} \ll c^2$, the following polynomial equation in $\delta\omega$.  
\begin{equation} \label{eq-alf-gr-dis}
    \delta\omega^3-X \delta\omega^2+Y=0,
\end{equation}
where $X= AB$ and $Y=AC$ with 
\begin{equation}\label{eq-alf-coeff}
    \begin{split}
        &A=\frac{2\Omega_r^2\sin^2 \lambda \cos^2\theta}{ \widetilde{\Omega}_{A\pm}\left(\cos^2\theta V_A^2-V_{A}^2-{4\Omega_r^2\cos^2 \lambda}/{k^2}\right)},\\
        &B=V_A^2-V_s^2,\\
        &C=V_N^2 c^2 k^2+\frac{1}{4}\frac{ V_{\rm{osc}}^2 \Omega_0^2 {\cal E}_0}{\hbar \widetilde{\Omega}_{A\pm}}.
    \end{split}
\end{equation}
 Equation \eqref{eq-alf-gr-dis} is a cubic polynomial in $\delta\omega$ with the discriminant of its derivative being positive. So, it must have one real root and two complex-conjugate roots. However, out of these three roots, we will consider only the complex root with a positive imaginary part to study the neutrino-driven Alfv{\'e}n wave instability. Thus, we numerically solve Eq. \eqref{eq-alf-gr-dis} for $\Im(\delta\omega)>0$ with the same set of parameters as for the magnetosonic wave instability (\textit{cf}. Figs. \ref{fig2-fsmag_b} and \ref{fig3-fsmag_n}), and show the results in Figs. \ref{fig4-gr-alfven} and \ref{fig5-gr-alfven-n0}.
\par 
 Figure \ref{fig4-gr-alfven} shows the profiles of growth rates for neutrino-driven fast [Subplot (a)] and slow [Subplot (b)] Alfv{\'e}n waves. These result from coupling with magnetosonic modes due to the Coriolis force. The solid curve depicts only the neutrino beam effect. The dashed curve includes both the neutrino beam and two-flavor oscillation effects. In both subplots, there is a singularity at $\theta\approx\pi/2$. At this point, the shear Alfv{\'e}n wave remains undefined. The instability curves behave differently in the two intervals: $0<\theta<\pi/2$ and $\pi/2<\theta<\pi$. For a constant magnetic field $B_0\sim10^7$ T and rotational frequency $\Omega_r\sim5\times10^3~\rm{s}^{-1}$, the growth rates for both $\gamma_+$ and $\gamma_-$ decrease with $\theta$ in $0<\theta<\pi/2$. In $\pi/2<\theta<\pi$, they increase, but only when we consider the neutrino beam's influence. When we include both the neutrino beam and two-flavor oscillations, $\gamma_+$ increases with $\theta$  in $0<\theta<\pi/2$. The $\gamma_-$ shows a decreasing trend and becomes reduced. In the interval $\pi/2<\theta<\pi$, $\gamma_+$ is reduced but still increases with $\theta$, while $\gamma_-$ is enhanced but continues to decrease. Thus, when two-flavor oscillations are present, the modified fast Alfv{\'e}n wave is likely unstable in $0<\theta<\pi/2$. Strong instability for the slow mode appears in $\pi/2<\theta<\pi$. However, with an enhanced magnetic field (indicated by the dash-dotted lines), the growth curves for both fast and slow modes exhibit similar features and higher growth rates.
 \par 
We have also studied the influences of the plasma number density $(n_0)$ and rotational angle on the instability profiles of the neutrino-driven shear Alfv{\'e}n waves modified by the coupling effect of neutrino-driven magnetosonic waves caused by the Coriolis force. Figure \ref{fig5-gr-alfven-n0} shows that the impact of the plasma number density over the entire domain of $\theta$ is not significant except close to $\theta=\pi/2$, where it has a peak or hike in the growth rate. However, as we reduce the rotational angle, a decrease in the growth rate for both the fast [Subplot (a)] and slow [Subplot (b)] Alfv{\'e}n waves is noted.    
\section{Discussion and conclusion} \label{sec-con}
This work presents a novel analysis: we generalize previous studies by identifying the significant coupling between neutrino-driven Alfv{\'e}n and magnetosonic wave modes in a rotating magnetized plasma, a modification not reported before. To summarize, we have studied the oblique propagation of neutrino-driven MHD waves, their coupling, and associated instabilities in a rotating magnetized plasma. This study also incorporates the impacts of a neutrino beam and two-flavor oscillations. Starting from a neutrino MHD model similar to \cite{haas2016neutrino}, with weak neutrino beam-plasma interactions, we have derived a general dispersion relation. This relation involves resonance interactions between the neutrino beam and waves, neutrino two-flavor oscillations, and the Coriolis force. We found that neutrino effects contribute to shear Alfv{\'e}n waves through the Coriolis force in rotating fluids. This force also couples the Alfv{\'e}n waves to oblique magnetosonic waves. The result is significant modifications to each mode, as well as to the associated instabilities, driven by the coupling between the other mode and the rotational frequency. Thus, the present work advances previous studies that focused only on neutrino-driven magnetosonic waves and instabilities \citep{chatterjee2023neutrino,haas2016neutrino}. The coupling between Alfv{\'e}n and magnetosonic waves can disappear when the axis of rotation coincides with the direction of the magnetic field or when one disregards fluid rotation.     
\par  
We have considered the model that uses non-idealized geometry (in which the propagation vector is oblique to the magnetic field and the angular velocity of the rotating fluid is in a plane) to perform a linear analysis of specific physical phenomena, whose existence and behavior are too complex to isolate in highly nonlinear, multidimensional simulations of realistic astrophysical scenarios. The mapping involves applying realistic values for key parameters, such as the neutrino beam density, neutrino streaming energy, plasma number density, and the magnetic field, to the simplified model, rather than reproducing the full complexity of the actual environment's geometry or dynamics. \\
Thus, the geometry and assumptions in this work represent a simplification of the complex, dynamic conditions found in CCSN and PNS environments. The model provides a theoretical framework by analyzing a simplified system and applying specific parameters (such as those of SN1987A) to illustrate potential instability growth rates, which can be used to understand the underlying physical mechanisms in these extreme astrophysical scenarios. \\
The model maps onto realistic CCSN/PNS conditions in the following ways:
\begin{itemize}
\item The present work uses realistic parameters for density, temperature, and magnetic field strengths pertinent to a Type II CCSN, demonstrating the relevance of the theoretical instabilities to these environments.
\item The model successfully isolates and studies the specific coupling mechanisms, such as the influence of the Coriolis force on shear Alfvén and Magnetosonic waves, and the effects of neutrino flavor oscillations on instability growth rates. Such an investigation helps theoretically confirm the existence of these effects in a controlled environment.
\item The results of the model (e.g., enhanced growth rates for MHD waves under specific conditions) can provide insights into potential mechanisms (like neutrino heating, the generation of turbulence, or magnetic field amplification) that could facilitate the actual supernova explosion, which current multidimensional simulations struggle to model perfectly from first principles.
\item The model's findings can contribute to a qualitative understanding of the interplay among rotation, magnetic fields, and neutrinos, which may help interpret results from more complex, full-scale numerical simulations. 
\end{itemize} 
\par 
We have numerically studied the growth rates of instabilities in modified fast and slow magnetosonic waves, as well as in fast and slow Alfvén waves, with parameters relevant to type II supernova explosions due to the core collapse of massive stars \cite{bethe1990supernova}. We observed that compared to Ref. \cite{chatterjee2023neutrino}, the instability growth rates of magnetosonic modes become enhanced due to coupling with the Alfvén waves caused by the Coriolis force. A novel feature appears in the slow magnetosonic mode at relatively low magnetic field strengths. In this situation, instead of showing minimum growth rates near $\theta=0$ and $\theta=\pi$ (parallel and antiparallel wave propagation, as in Ref. \cite{chatterjee2023neutrino}), the slow magnetosonic mode now exhibits maximum growth rates. Also, unlike Ref. \cite{chatterjee2023neutrino}, the growth rate for the slow magnetosonic mode decreases in the interval $0<\theta<\pi/2$, while it increases in $\pi/2<\theta<\pi$.
On the other hand, when the magnetic field strength is relatively high, the features remain similar to Ref. \cite{chatterjee2023neutrino}. However, the growth rates become slightly enhanced for both fast and slow magnetosonic waves. Further, a small increase in the plasma number density $n_0$ can significantly enhance the growth rate for the fast magnetosonic mode. This impact on the slow-mode growth rate is almost insignificant. In contrast, slightly decreasing the rotational angle $\lambda$ can significantly reduce the slow mode's growth rate, with little effect on the fast mode.      
\par
 We have revealed new neutrino-driven shear Alfv{\'e}n (both fast and slow) modes to exist due to the influence of the Coriolis force. Notably, previous authors neither reported any neutrino contribution to Alfv{\'e}n waves nor their coupling with magnetosonic waves. Building on this distinction, we have shown that modified fast and slow shear Alfv{\'e}n waves remain unstable over the entire domain of $\theta$: $0<\theta<\pi$ except at $\theta=\pi/2$, where the wave is undefined. Furthermore, we observed that the growth rates for shear Alfv{\'e}n waves in two regimes of $\theta$, namely $0<\theta<\pi/2$ and $\pi/2<\theta<\pi$, exhibit almost opposite characteristics. Such behaviors indicate that, depending on the angle of propagation, the growth rates for fast and slow modes can be either enhanced or reduced by the influence of neutrino two-flavor oscillations. At the same time, the magnetic field has a noticeable impact on the Alfv{\'e}nic growth rate, a reduction in the rotational angle, and a plasma density enhancement result in a decrease in the growth rate over the domain of the angle of propagation except near $\theta=\pi/2$.   
\par 
We mention that typical values of the magnetic field and plasma number density (relevant to the protoneutron star surface) considered in the 3D MHD simulations of core-collapse supernovae \cite{nakamura2025} are $B_0\sim 10^6 T$ (initial) and $n_0\sim10^{34}~\rm{m}^{-3}$). In the present work, we have considered parameter values close to these. We have also shown how the characteristics of the growth rates for magnetosonic and Alfv{\'e}n waves change due to small changes in these parameters. We have observed that retaining the plasma number density at $n_0\sim10^{34}~\rm{m}^{-3}$, the qualitative features of the growth rates for fast and slow magnetosonic waves in the entire domain of the obliqueness of the propagation angle remain the same if the magnetic field strength $(B_0)$  varies in the range $10^6\lesssim B_0<10^7$ T, as well as at a lower strength  $B_0\sim10^5$ T. However, as the magnetic field increases within the domain: $10^6\lesssim B_0<10^7$ T, the growth rate of the fast magnetosonic mode decreases, while that of the slow mode increases. The qualitative features are significantly changed for the magnetic field satisfying $B_0\gtrsim10^7$ T. Not only, instead of the inverted bell shaped and double humped curves (except for the slow mode when the neutrino beam, two-flavor oscillations, and the coupling with the Alfv{\'e}n effects are present), there appeared bell-shaped and $\gamma$-shaped curves for the fast and slow modes, in contrast to previous case, the instability growth rate is maximized near the perpendicular propagation for the fast mode, and near parallel and antiparallel propagations for the slow mode. As in the previous regime, $10^6\lesssim B_0<10^7$ T, the growth rate for the fast mode decreases, while that for the slow mode increases with increasing $B_0$ in $B_0\sim10^7-10^8$ T. However, for $B_0\gtrsim10^8$ T, although the growth rate for the fast mode remains decreasing with increasing values of $B_0$, the same for the slow mode tends to remain unchanged in the entire domain of the obliqueness angle $\theta$. Thus, the results for instability growth rates are robust across a wide range of magnetic field strengths. Typically, for $\delta m^2 c^4\sim3\times10^{-5}~\rm(ev)^2$, ${\cal E}_0\sim 1~{\rm Mev}$, we have $\omega_0\sim2.3\times10^4~{\rm s}^{-1}$. So, for a neutrino oscillation mixing angle, $\theta_0\sim3\degree$, one obtains $\Omega_0\sim2.3\times10^3~{\rm s}^{-1}$. Thus, for maximum growth rates of magnetosonic waves, i.e., for $\gamma_{\pm}/\Omega_0\sim3-5\times10^{-3}$, the instability times vary in the range, $0.09-0.14$ s. Thus, we have instability times within the predicted time of the neutrino-driven explosion ($0.3$ s after bounce) reported in the simulation \cite{nakamura2025}. 
On the other hand, the instability growth rates for Alfv{\'e}n waves appear to be low $(\gamma_{\pm}\sim10^{-6})~{\rm s}^{-1}$ compared to magnetosonic waves.
For maximum growth rates with $\gamma_{\pm}/\Omega_0\sim2-4.5\times10^{-6}$, the instability times vary in the range $143-333$  s, i.e., longer than the predicted time for magnetosonic waves. While shear Alfv{\'e}n waves may be phenomenal in other circumstances, in the context of neutrino-driven instabilities, such as those in core-collapse supernovae explosions, the compressional natures of magnetosonic waves provide a superior mechanism for energy extraction from the neutrino beam, giving higher instability growth rates.
We, however, note that the instability growth rates can be higher or instability times can be shorter for both magnetosonic and Alfv{\'e}n waves with a higher neutrino fluid or plasma density. This growing instability arises because higher density enhances resonant coupling between neutrinos and plasmas via the weak force. This stronger interaction leads to a more efficient transfer of energy from the neutrino beam to the MHD waves, thereby accelerating the growth of instability. 
\par 
We have noted that the maximum growth rates for magnetosonic waves typically vary within $\gamma_{\pm}/\Omega_0\sim3-5\times10^{-3}$, i.e., $\gamma_{\pm}\sim7-11~{\rm s}^{-1}$ for $\Omega_0\sim2.3\times10^3~{\rm s}^{-1}$. Thus, the corresponding growth length, $L=c/\gamma_{\pm}$, varies as $(3-4)\times10^4$ km. Although such lengths are much larger than the typical length scale of the stalled shock radius ($100-200$ km), they are, however, consistent with the shock radius (more than several thousands of kilometers) observed at the end of the 3D MHD simulations of CCSN models that exhibit shock revival \cite{nakamura2025}. The growth length corresponding to Alfv{\'e}n waves becomes even larger than for magnetosonic waves because of lower growth rates.
In typical multidimensional CCSN models, successful explosions occur by heating neutrinos assisted by hydrodynamic instabilities such as convective motion, thereby increasing the retention time of the shocked material within the gain region \cite{foglizzo2006} and the Standing Accretion Shock Instability \cite{foglizzo2007}. The MHD instabilities, associated with coupled Alfv{\'e}n and magnetosonic waves, reported here, can enhance the efficiency of neutrino heating by expanding the heating region and allowing accreting matter to spend more time in the gain region where neutrinos deposit energy into the matter. Such an increase in the dwell time of shocked material leads to a more efficient energy transfer from neutrinos to plasmas. This enhanced neutrino-plasma coupling can increase the likelihood of reviving the stalled shock, enabling the explosion to proceed outward.    
\par
To conclude, the neutrino-driven MHD wave coupling and instabilities reported here could play a vital, synergistic role in the explosion of core-collapse supernovae. Since the coupled Alfv{\'e}n and magnetosonic waves resonantly interact with streaming neutrino beams and two neutrino-flavor oscillations, they extract energy from the neutrino beam to drive MHD instabilities that can revive stalled shocks and accelerate the supernova explosion. Depending on the angle of propagation relative to an external magnetic field, the instability growth rate can further be enhanced, and the wave energy can then be amplified by this NMHD instability, leading to a relatively faster blow-up of the star. We also mention that the neutrino two-flavor oscillations modulate the weak interaction force between the streaming neutrino beam and background plasmas. So they directly influence the growth rates of instability. Resonant interactions occur (or instability growth is maximized) when the MHD wave frequency matches the frequencies associated with the streaming beam and neutrino flavor oscillation. The latter plays a role of additional driving or beating frequency, triggering resonant energy transfer from the neutrino beam to plasmas during core-collapse supernovae. Although direct observations of the effect of neutrino flavor oscillations on MHD wave instability remain challenging, their macroscopic consequences are likely to be observable in the foreseeable future.
Furthermore, supernova explosion data can uniquely constrain neutrino mass-squared differences through phenomena such as signal time delays, spectral modifications, and the determination of the neutrino mass hierarchy \cite{scholberg2018}. We have restricted our investigation to the ideal NMHD plasma model. The inclusion of finite conductivity and the Hall current effect can be a fruitful avenue for future studies. 
\section*{Acknowledgements}
One of us, J. Turi wishes to thank the Council of Scientific and Industrial Research (CSIR) for a Senior Research Fellowship (SRF) with reference number 09/202(0115)/2020-EMR-I.
\section*{Author Contributions}
\textbf{Jyoti Turi:} Formal analysis (equal); Investigation (equal); Methodology (equal); Software (equal); Visualization (equal); Validation(equal) Writing–original draft (equal).\textbf{ Amar P. Misra:} Conceptualization
(lead); Formal analysis (equal); Investigation (lead); Methodology (lead); Software (equal); Visualization (equal); Supervision (lead); Validation (equal); Writing-review \& editing (lead).
\section*{Data Availability}
All data that support the findings of this study are included within the article and also available in \cite{nakamura2025}.


\bibliographystyle{mnras}
\bibliography{ref1} 








\bsp	
\label{lastpage}
\end{document}